\documentclass
[%
reprint,
superscriptaddress,
showpacs,
amsmath,amssymb,
aps,
pre,
floatfix,
]
{revtex4-2}

\usepackage{xcolor}
\usepackage[utf8]{inputenc}
\usepackage[english]{babel}
\usepackage{graphicx}
\usepackage{comment}
\usepackage{siunitx}
\usepackage{multirow}

\begin{document}


\title{Noise-induced Extreme Events in Hodgkin-Huxley Neural Networks}
\author{Bruno R. R. Boaretto}
\email{bruno.boaretto@unifesp.br}
\affiliation{Institute of Science and Technology, Universidade Federal de S\~ao Paulo, 12247-014, S\~ao Jos\'e dos Campos, S\~ao Paulo, Brazil}
\affiliation{Department of Physics, Universitat Politecnica de Catalunya, 08222, Terrassa, Barcelona, Spain}
\author{Elbert E. N. Macau} 
\email{elbert.macau@unifesp.br}
\affiliation{Institute of Science and Technology, Universidade Federal de S\~ao Paulo, 12247-014, S\~ao Jos\'e dos Campos, S\~ao Paulo, Brazil}
\author{Cristina Masoller}%
 \email{cristina.masoller@upc.edu}
\affiliation{Department of Physics, Universitat Politecnica de Catalunya, 08222, Terrassa, Barcelona, Spain}

\begin{abstract}

Extreme events are rare, large-scale deviations from typical system behavior that can occur in nonlinear dynamical systems. In this study, we explore the emergence of extreme events within a network of identical stochastic Hodgkin-Huxley neurons with mean-field coupling. The neurons are exposed to uncorrelated noise, which introduces stochastic electrical fluctuations that influence their spiking activity. Analyzing the variations in the amplitude of the mean field, we observe a smooth transition from small-amplitude, out-of-sync activity to synchronized spiking activity as the coupling parameter increases, while an abrupt transition occurs with increasing noise intensity. However, beyond a certain threshold, the coupling abruptly suppresses the spiking activity of the network. Our analysis reveals that the influence of noise combined with neuronal coupling near the abrupt transitions can trigger cascades of synchronized spiking activity, identified as extreme events. The analysis of the entropy of the mean field allows us to detect the parameter region where these events occur. We characterize the statistics of these events and find that, as the network size increases, the parameter range where they occur decreases significantly. Our findings shed light on the mechanisms driving extreme events in neural networks and how noise and neural coupling shape collective behavior.
\end{abstract}
\maketitle

\section{Introduction}

Extreme events are rare and sudden large-scale deviations from typical behavior \cite{chowdhury2022extreme,kantz2006,hobsbawm2020age}. These events manifest across a variety of natural, technological, and social systems, including earthquakes \cite{bak1989earthquakes}, tsunamis \cite{field2012managing}, climatic phenomena  \cite{herring2015explaining,kurths2019}, rogue waves in oceans \cite{ocean} and in optical systems \cite{solli2007,zamora2013rogue}, large-scale blackouts in power grids \cite{kurths2023}, neuronal avalanches \cite{beggs2003neuronal}, epileptic seizures \cite{lehnertz2006epilepsy,frolov2019statistical}, and market crashes \cite{sornette2009stock}, to name a few.

In excitable systems, significant progress has been made in understanding the mechanisms that lead to extreme events. In FitzHugh-Nagumo neurons, a possible mechanism involves recruitment dynamics within coupled systems: a small subset of units becomes excited, subsequently recruiting others through diffusive coupling, until a large portion or even the whole
system exhibits synchronized excitation, manifesting as an extreme event \cite{ansmann2013extreme,feudel2,ansmann2016self}. Evidence of such dynamics has also been found in Hindmarsh-Rose bursting neurons that interact through different types of coupling configurations, such as chemical synaptic and gap junctional-type diffusive coupling \cite{mishra2018dragon}. 

Synchronized activity has been observed in simulated and real neural systems \cite{bressloff2000dynamics,batista2009bursting,hao2011single,angulo2017death,andreev2018coherence,tibau2018analysis,boaretto2021role,hernandez2021noise,faci2023dynamical,ram2024spatiotemporal,farrera2024neuron,efimova2024spiking}. For instance, synchronous neural activity in a particular brain region of canaries (where sensorimotor integration occurs) emerges in response to auditory playback of the bird's own song \cite{ana_2022}. Well‐defined oscillations in the local field potentials were recorded, which were locked to song rhythm. 
However, noise plays a crucial role in neural activity, amplifying or triggering resonant dynamics. For example, noise-driven networks of FitzHugh-Nagumo oscillators transition from irregular, intermittent synchronized events to more regular occurrences as noise levels increase \cite{zaks2005noise}. As another example, noise can enhance the ability of FitzHugh-Nagumo neurons to encode weak signals \cite{masoliver2020neuronal}. However, how noise interacts with intrinsic neuronal properties and network topology to produce extreme events is still far from being understood.

A recent study has explored the emergence of phase synchronization in neuronal networks driven by Poissonian inputs and coupled through chemical synapses, highlighting the intricate balance between external stimulation and coupling currents \cite{boaretto2023phase}. 
With this motivation, in this work we study the effects of uncorrelated noise when the neuronal network has global mean field coupling. We use the classic Hodgkin-Huxley model \cite{hodgkin1952quantitative} that captures the spiking behavior of neurons through a stable limit cycle when excited above a threshold \cite{ermentrout2010mathematical,izhikevich2007dynamical}. By incorporating stochasticity, we simulate neurons in a noisy environment, where random electrical fluctuations influence their behavior in the absence of any constant bias current. For low levels of noise, the neuron displays subthreshold oscillations around an equilibrium point, while at sufficiently high noise levels, the neurons transition to irregular spiking activity.

Here, the neurons are coupled through diffusive electrical interactions, with each neuron in the network subjected to uncorrelated noise. This combination of noise and coupling can lead to complex macroscopic behaviors, including incoherence, intermittency, synchrony, and avalanche-like phenomena, here referred to as extreme events \cite{beggs2003neuronal}. These events emerge as a cascade effect, where a significant portion of neurons rapidly exhibits spiking activity. Synchronization transitions are also found, which resemble those occurring in other models of networks of stochastic spiking neurons~\cite{roque_srep_2016}.

We detect the extreme events by analyzing variations in the amplitude of the mean field and its entropy, and identify the parameter region where they are most likely to occur. Our results show that the time intervals between extreme events follow an exponential distribution. We also show that, as the network size increases, the region where extreme events occur decreases substantially. 

The paper is organized as follows:  Sec. \ref{sec:neuronal_model} presents the model, Sec. \ref{sec:global} presents the results, and Sec. \ref{sec:conc} presents the discussion and our conclusions.

\section{Model}\label{sec:neuronal_model}

To simulate the spiking neuronal dynamics, we consider the Hodgkin-Huxley (HH) model \cite{hodgkin1952quantitative}, which describes how the time evolution of the membrane potential of the neuron measured in \SI{}{\milli\volt} (millivolts) is related to the variations of two voltage-gated channels associated with the ion concentrations of potassium ($\mathrm K ^+$) and sodium ($\mathrm{Na}^+$), as well as a leakage channel associated with the passive variations (non-gated channels) \cite{ermentrout2010mathematical}. To study the collective behavior of $N$ identical neurons with global, mean-field coupling, the membrane potential of each neuron, $V_i$ with $i=1,\cdots, N$, is described by
\begin{eqnarray}\label{eq:N_hh}
    C_\mathrm M \frac{dV_i}{dt}&=& -g_\mathrm K n_i^4(V_i-E_\mathrm K) - g_\mathrm{Na}m_i^3h_i(V_i-E_\mathrm{Na}) \nonumber \\ && - g_\mathrm{\ell}(V_i-E_\mathrm \ell) + D\xi_i(t) + I_{i,\mathrm{coup}}.
\end{eqnarray}
Here, $C_\mathrm M$ is the capacitance of the cell membrane, $g_\mathrm K$, $g_\mathrm{Na}$ and $g_\mathrm \ell$ are the maximum conductances, and $E_\mathrm K$, $E_\mathrm{Na}$, and $E_\mathrm \ell$ are the reversal potential of each ionic current. $\xi_i(t)$ represents a Gaussian noise with zero mean and standard deviation one, and $D$ is a parameter that controls the noise magnitude. 

The variables $n_i$ and $m_i$ are related to activating the potassium and sodium ionic currents, respectively, and $h_i$ is the inactivation of the sodium current. Their time evolution is described by: 
\begin{eqnarray}    
\frac{dn_i}{dt} &=& \alpha_{n,i}(1-n_i) - \beta_{n,i} n_i, \label{eq:n}\\
\frac{dm_i}{dt} &=& \alpha_{m,i}(1-m_i) - \beta_{m,i} m_i\\
\frac{dh_i}{dt} &=& \alpha_{h,i}(1-h_i) - \beta_{h,i} h_i, \label{eq:h}
\end{eqnarray}
where  $\alpha$ and $\beta$ are functions that depend on $v_i=V_i/{\rm mV}$:
\begin{eqnarray}\label{eq:alpha_1}
\alpha_{n,i}&=&\frac{0.01(v_i+55)}{(1-\exp[-(v_i+55)/10])},\\
\alpha_{m,i}&=&\frac{0.1(v_i+40)}{(1-\exp[-(v_i+40)/10])},\\
\alpha_{h,i}&=&0.07\exp[-(v_i+65)/20],\\
\beta_{n,i}&=&0.125\exp[-(v_i+65)/80],\\
\beta_{m,i}&=&4\exp[-(v_i+65)/18],\\
\beta_{h,i}&=&\frac{1}{(1+\exp[-(v_i+35)/10])}.\label{eq:alpha_2}
\end{eqnarray} 

$I_{i,\mathrm{coup}}$ represents the current due to coupling,
\begin{equation}\label{eq:coupling}
I_{i,\mathrm{coup}} = \frac{\varepsilon}{N}\sum_{j=1}^N(V_j - V_i) = \varepsilon(\overline V - V_i),   
\end{equation}
where $\varepsilon$ is the coupling parameter, $V_i$ ($V_j$) is the membrane potential of the postsynaptic (presynaptic) cell, and $\overline V$ is 
the mean field of the network,
\begin{equation}
\overline V= \frac{1}{N}\sum_{i=1}^NV_i.
\end{equation}

One of the main characteristics of the HH model is the existence of a sub-critical Andronov-Hopf bifurcation as the external current increases \cite{ermentrout2010mathematical,izhikevich2007dynamical}. Hence, when the neuron is stimulated above a threshold, the stable equilibrium point loses stability leading to the emergence of a limit cycle that corresponds to periodic spiking activity. In the absence of stimulation, the neuron rests at the equilibrium point at $V^*\approx -65$ $\SI{}{\milli\volt}$. 
\begin{figure}[tb!] 
    \centering
    \includegraphics[width=.75\columnwidth]{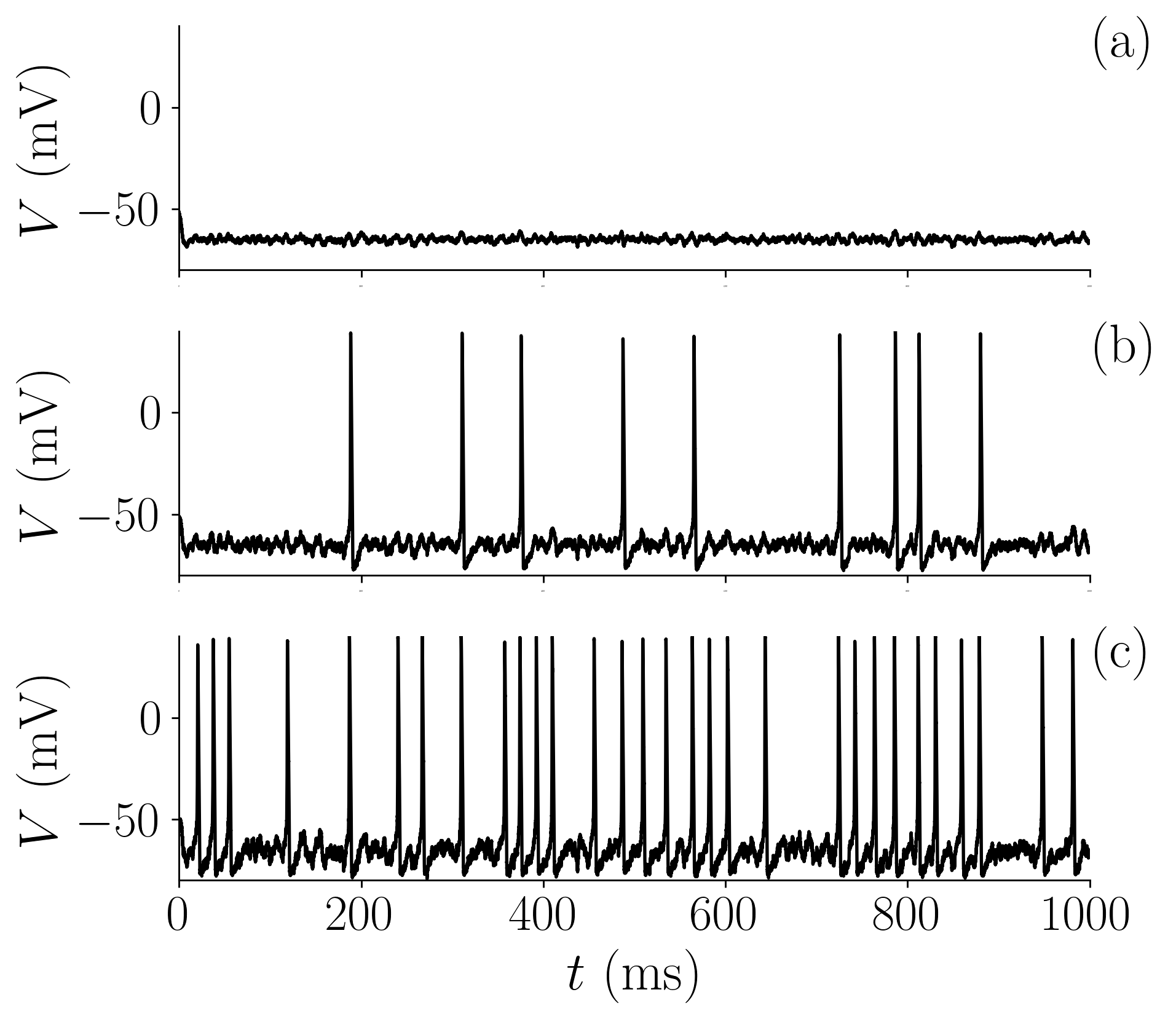}
    \caption{Dynamics of an isolated HH neuron ($I_{i,\mathrm{coup}} = 0$) considering the parameters of Table \ref{tab:tablehh}, for (a) $D=1$, (b) $D=2$, and (c) $D=3$ ($\SI{}{\micro\ampere/\centi\meter^2}$).}
    \label{fig:hh_dyn}
\end{figure}

\begin{table}[tb!] 
\setlength{\tabcolsep}{7pt}
\centering
\caption{Parameters used in the simulations \cite{ermentrout2010mathematical}.}
\label{tab:tablehh}
\begin{tabular}{l l r}
    \hline \hline
     {Membrane capacitance
     ($\SI{}{\micro\farad/\centi\meter^{2}}$)} & $C_\mathrm M$ & $1$ \\ \hline
    \multirow{3}{*}{Maximum conductances
($\SI{}{\milli\siemens/\centi\meter^{2}}$)}  
&$g_\mathrm{Na}$&$120$  \\ 
&$g_\mathrm K$&$36$  \\ 
&$g_\mathrm \ell$&$0.3$\\
\hline
    \multirow{3}{*}{Resting potentials
 (\SI{}{\milli\volt})}  & $E_\mathrm{Na}$&$50$ \\  & $E_\mathrm K$&$-77$ \\   & $E_\mathrm \ell$&$-54.4$    \\
    \hline
\end{tabular}
\end{table}

{The model equations were integrated from random initial conditions, with each neuron at a random point in the phase space. Specifically, $V(0)$ was chosen randomly in $[-80, 0]$, while $n(0)$, $m(0)$, and $h(0)$ were chosen randomly in $[0, 1]$. The integration was performed using the Heun method for stochastic systems \cite{garcia2012noise}, with a time step $\Delta t = 0.01$ $\SI{}{\milli\second}$.} The parameters used are listed in Table~\ref{tab:tablehh}. To disregard transient effects we discarded the first $10^4$ $\SI{}{\milli\second}$. The dynamics of an isolated neuron ($I_{i,\mathrm{coup}} = 0$) is illustrated in Fig. \ref{fig:hh_dyn} for different noise strengths: In panel (a), the noise is not strong enough to generate spikes, while in (b) and (c), the noise induces irregular spiking activity, and we observe that the stronger the noise, the greater is the number of spikes. 

\begin{figure*}[tb!]
    \centering
    \includegraphics[width=.95\linewidth]{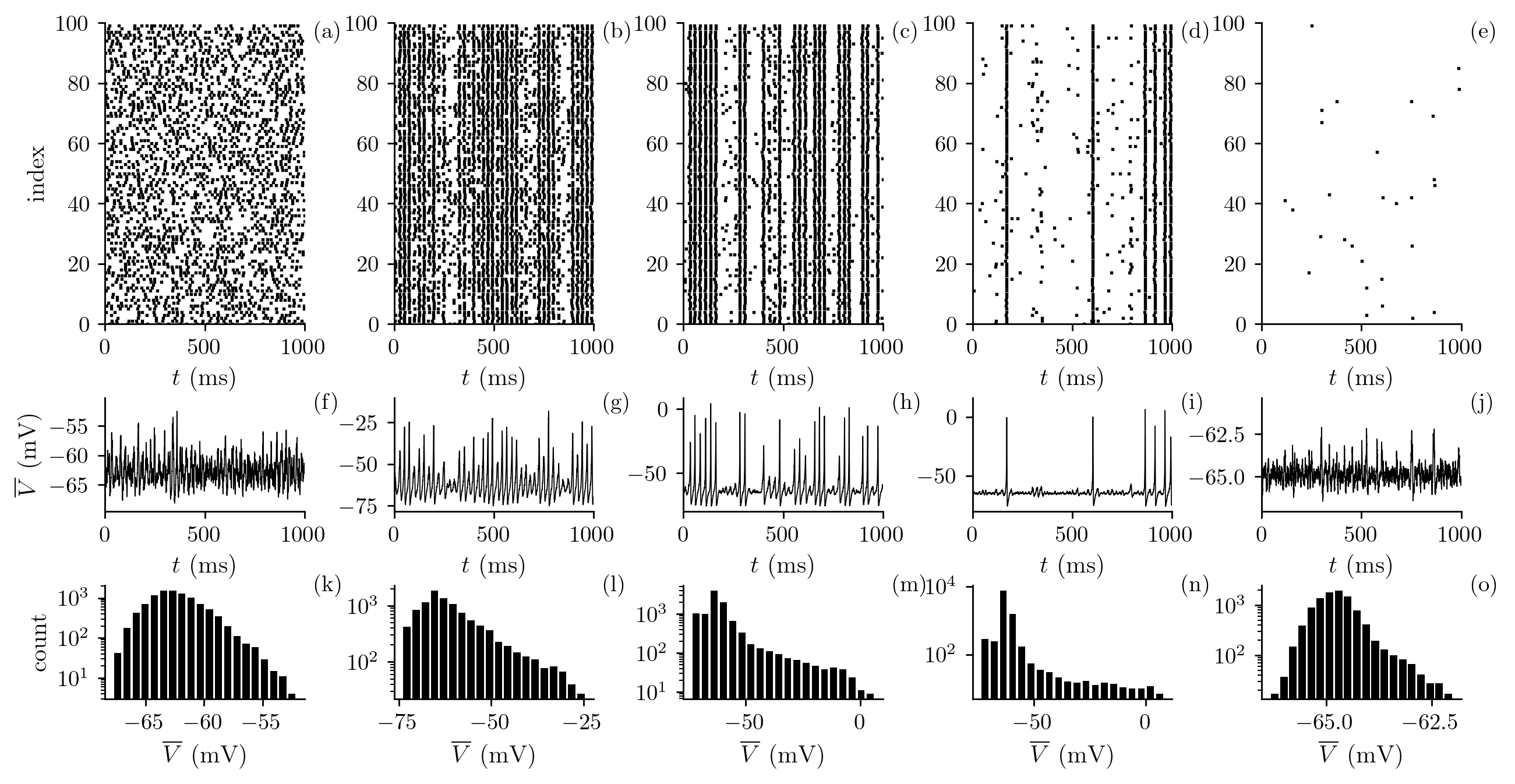}
    \caption{Raster plot of a network of 100 neurons, the time evolution of the mean field, $\overline{V}$, and histograms for increasing coupling strength at fixed noise intensity, $D=3$ ($\SI{}{\micro\ampere/\centi\meter^2}$). Panels (a, f, k) $\varepsilon=0.1$: incoherent macroscopic activity. Panels (b, g, l) $\varepsilon=0.3$: onset of phase synchronization. Panels (c, h, m) $\varepsilon=0.5$ $\SI{}{\milli\siemens/\centi\meter^{2}}$: pronounced spiking activity. Panels (d, i, n) $\varepsilon=0.8$ $\SI{}{\milli\siemens/\centi\meter^{2}}$: reduced spiking frequency. Panels (e, j, o) $\varepsilon=1.0$ $\SI{}{\milli\siemens/\centi\meter^{2}}$: almost suppression of the spiking activity.}
    \label{fig:LFP_add_eps}
\end{figure*}

\section{Results}\label{sec:global}

We analyze the dynamics of $N=100$ coupled neurons. Figure \ref{fig:LFP_add_eps} shows the raster plot (RP), the evolution of mean field $\overline{V}$, and mean field histograms, for increasing coupling strength, $\varepsilon$, and fixed noise intensity. The RP indicates the spiking times of individual neurons, where a spike is evaluated when $V_i$ crosses $-20$ $\SI{}{\milli\volt}$. In panels (a, f, k) ($\varepsilon=0.1$), despite high electrical activity driven by noise, the coupling current is too weak to induce phase synchronization. Consequently, $\overline{V}$ oscillates incoherently around the neuron's equilibrium state. As the coupling increases (panels b, g, l), the synaptic current induces larger oscillations of the mean field and the emergence of vertical structures in the raster plot, indicating increased synchronization. For higher coupling (panels c, h, m), the synchronization is more pronounced, with a clearer vertical alignment of spikes in the RP and well-defined mean field spiking activity. However, with a further increase of the coupling strength (panels d, i, n), this spiking activity becomes less frequent, and for high enough coupling (panels e, j, o), spiking activity disappears, and electrical activity is suppressed due to too strong coupling. The histograms of the mean field, shown in panels (k)–(o), illustrate how the shape of the distribution changes with increasing $\varepsilon$. 

These variations reflect the transitions of the electrical activity in the network. The damping of electrical activity observed with increasing coupling ($\varepsilon$) arises from the homogenization of the membrane potentials across the network. As $\varepsilon$ grows, the coupling term $\varepsilon (\overline{V} - V_i)$ increasingly drives individual neuron dynamics toward the mean field ($\overline{V}$), reducing differences between neurons. This suppresses the variability necessary for spiking activity, effectively diminishing oscillations and leading to a state where the network exhibits reduced or no spiking. This transition that abruptly suppresses the network spiking activity when the coupling increases above a critical value resembles the phenomena of amplitude and oscillation death \cite{koseska2013oscillation}. Additionally, this behavior aligns with the concept of over-synchronization, where excessive coupling eliminates the diversity of individual dynamics necessary for sustained collective oscillations \cite{pecora1997fundamentals}.  

\begin{figure*}[tb!] 
    \centering
    \includegraphics[width=0.95\linewidth]{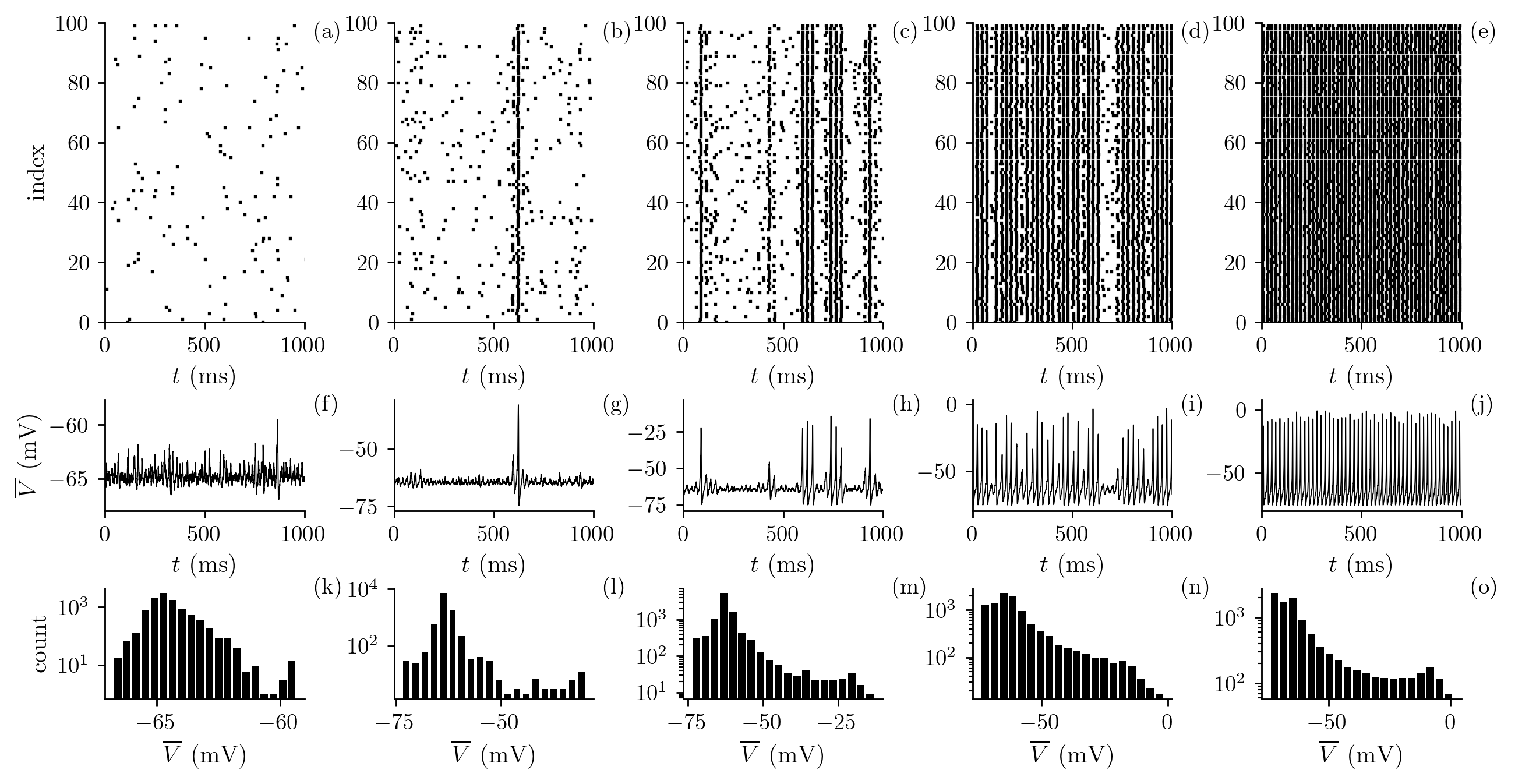}
    \caption{Raster plot of a network of 100 neurons and its mean field for increasing noise intensity at fixed coupling strength, $\varepsilon = 0.6$ $\SI{}{\milli\siemens/\centi\meter^{2}}$. (a, f, k) For $D=2.4$ ($\SI{}{\micro\ampere/\centi\meter^2}$), the network exhibits weak spiking activity, and its mean field oscillates around the equilibrium. (b, g, l) For $D=2.6$ ($\SI{}{\micro\ampere/\centi\meter^2}$), sporadic localized events emerge, where the majority of neurons spike simultaneously. As the noise intensity increases [in (c, h, m) $D=2.8$ {($\SI{}{\micro\ampere/\centi\meter^2}$)}; in (d, i, n) $D=3.0$ ($\SI{}{\micro\ampere/\centi\meter^2}$)], these events become more pronounced. For strong noise [in (e, j, o) $D=4.0$ ($\SI{}{\micro\ampere/\centi\meter^2}$)], the network exhibits sustained synchronized spiking activity.} 
    \label{fig:LFP_add_D}
\end{figure*}

Figure \ref{fig:LFP_add_D} shows the network dynamics for increasing noise intensities at a fixed coupling strength. For weak noise, panels (a, f, k), the mean field oscillates around the equilibrium point due to minimal electrical activity in the network. As noise intensity increases, panels (b, g, l), the spiking activity of individual neurons grows, and the interplay between noise and coupling allows for the emergence of sporadic, localized events where the neurons spike simultaneously before quickly returning to equilibrium; in the following, we will refer to these short, avalanche-like events as extreme events. The sudden occurrence of these events represents a tail in the histogram of the mean field. As the noise strength increases further, these events become longer and more frequent, as illustrated in panels (c, h, m), (d, i, n), and (e, j, o). 

\begin{figure}[tb!] 
    \centering
    \includegraphics[width=0.75\linewidth]{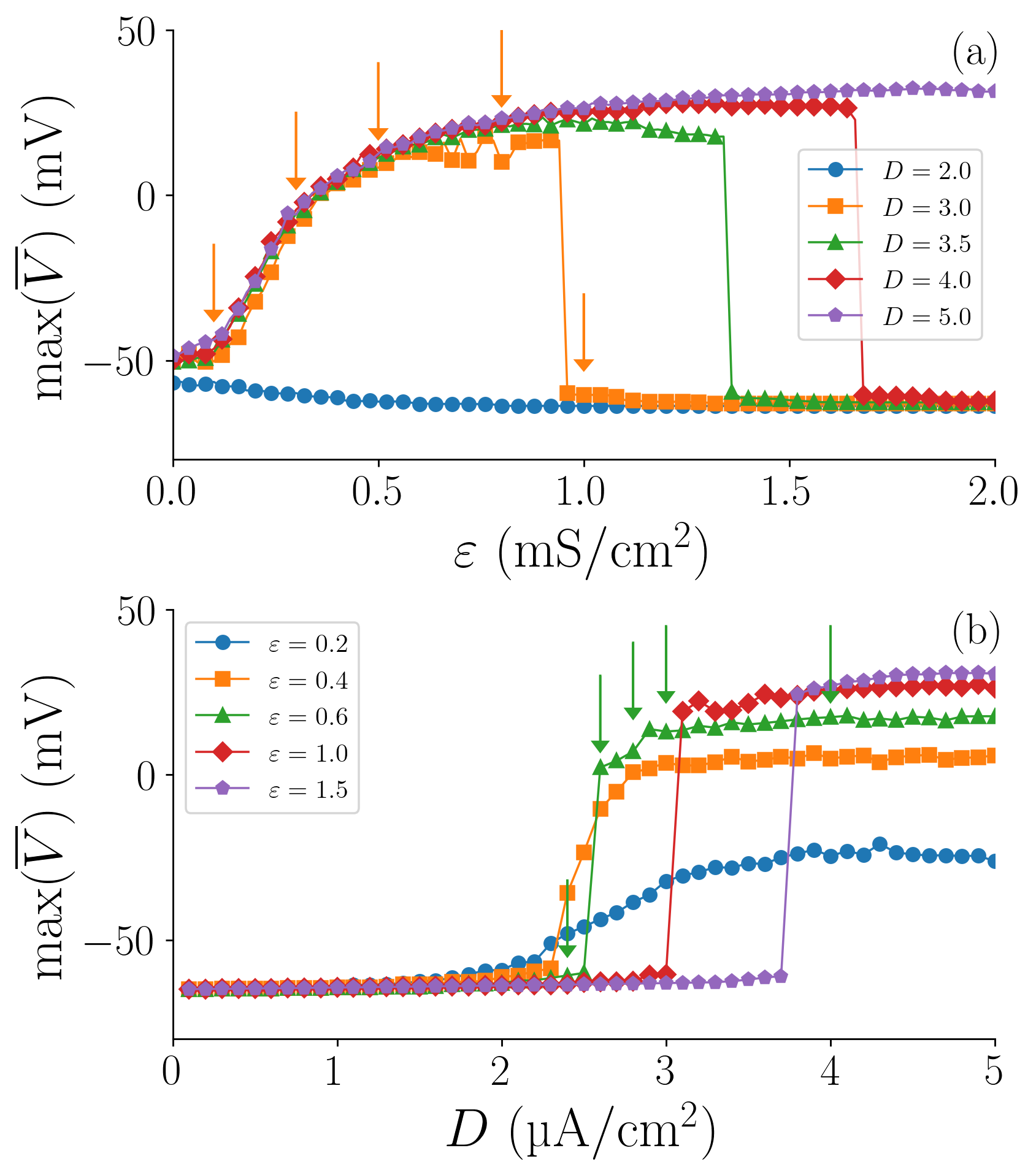}
    \caption{Maximum mean field amplitude, $\mathrm{max}(\overline{V})$, as a function of coupling strength $\varepsilon$ (a), and noise intensity $D$ (b). Panel (a) shows a gradual increase of $\mathrm{max}(\overline{V})$ with increasing $\varepsilon$, followed by an abrupt decrease at high $\varepsilon$. Panel (b) reveals a gradual increase of $\mathrm{max}(\overline{V})$ for weak coupling that becomes abrupt, at higher coupling, as noise intensity $D$ grows. The orange and green arrows indicate the values used in Figs. \ref{fig:LFP_add_eps} and \ref{fig:LFP_add_D}, respectively.}
    \label{fig:amp_analysis}
\end{figure}

Figure \ref{fig:amp_analysis} shows $\mathrm{max}(\overline{V})$ as a function of (a) the coupling strength, $\varepsilon$, and (b) the noise intensity, $D$. In Fig. \ref{fig:amp_analysis}(a), for sufficiently low values of $D$ (blue dots), $\mathrm{max}(\overline{V})$ remains small in all the range of coupling strengths studied, indicating the absence of significant collective behavior. As $D$ increases, a gradual transition occurs, where $\mathrm{max}(\overline{V})$ shifts from low to high values. This behavior suggests a synchronization transition similar to the one observed in networks of Kuramoto phase oscillators and various neuronal networks \cite{kuramoto1975self,boaretto2023phase}. In contrast, for larger values of $\varepsilon$, an abrupt change is observed: $\mathrm{max}(\overline{V})$ abruptly drops to low values due to the suppressive effect of the coupling discussed earlier. If we observe the panel (e) of Fig. \ref{fig:LFP_add_eps}, this drop in the amplitude of the mean field is associated with a suppression of the network electrical activity due to high coupling values. Higher noise intensities shift the critical coupling needed for this abrupt transition to higher values.  

In Fig. \ref{fig:amp_analysis}(b), for weak coupling strength, we observe only a small increase of $\mathrm{max}(\overline{V})$ with the noise strength. However, for sufficiently strong coupling, when the noise strength increases, at a threshold value of the noise strength $\mathrm{max}(\overline{V})$ undergoes an abrupt transition to much higher values. This indicates that, the larger the noise intensity, the larger the coupling strengths needed to achieve high-amplitude oscillations of the mean field. All abrupt transitions require that a significant fraction of the network spikes together. Consequently, the exact location of the transition varies with the initial conditions and the duration of the simulation, although the overall dynamics remains consistent. For future work, it will be interesting to test the phenomena of bistability and hysteresis. 

In addition, a similar abrupt transition induced by changes in noise intensity has been reported in globally coupled stochastic FitzHugh-Nagumo neuron arrays, where it is described as a ``canard explosion'' \cite{zaks2005noise}. This phenomenon occurs when system trajectories remain close to the repelling branch of a slow manifold before diverging, causing a sudden increase in oscillation amplitude \cite{rotstein2012canard}. In coupled excitable systems, interactions between units can synchronize this behavior, amplifying it across the network and leading to collective bursts. Such transitions are often associated with a subcritical Hopf bifurcation, where the system shifts from a stable equilibrium directly to large-amplitude oscillations as a critical threshold is crossed, further emphasizing the abruptness of the transition \cite{rotstein2012canard}.

\begin{figure}[tb!] 
    \centering
    \includegraphics[width=.98\columnwidth]{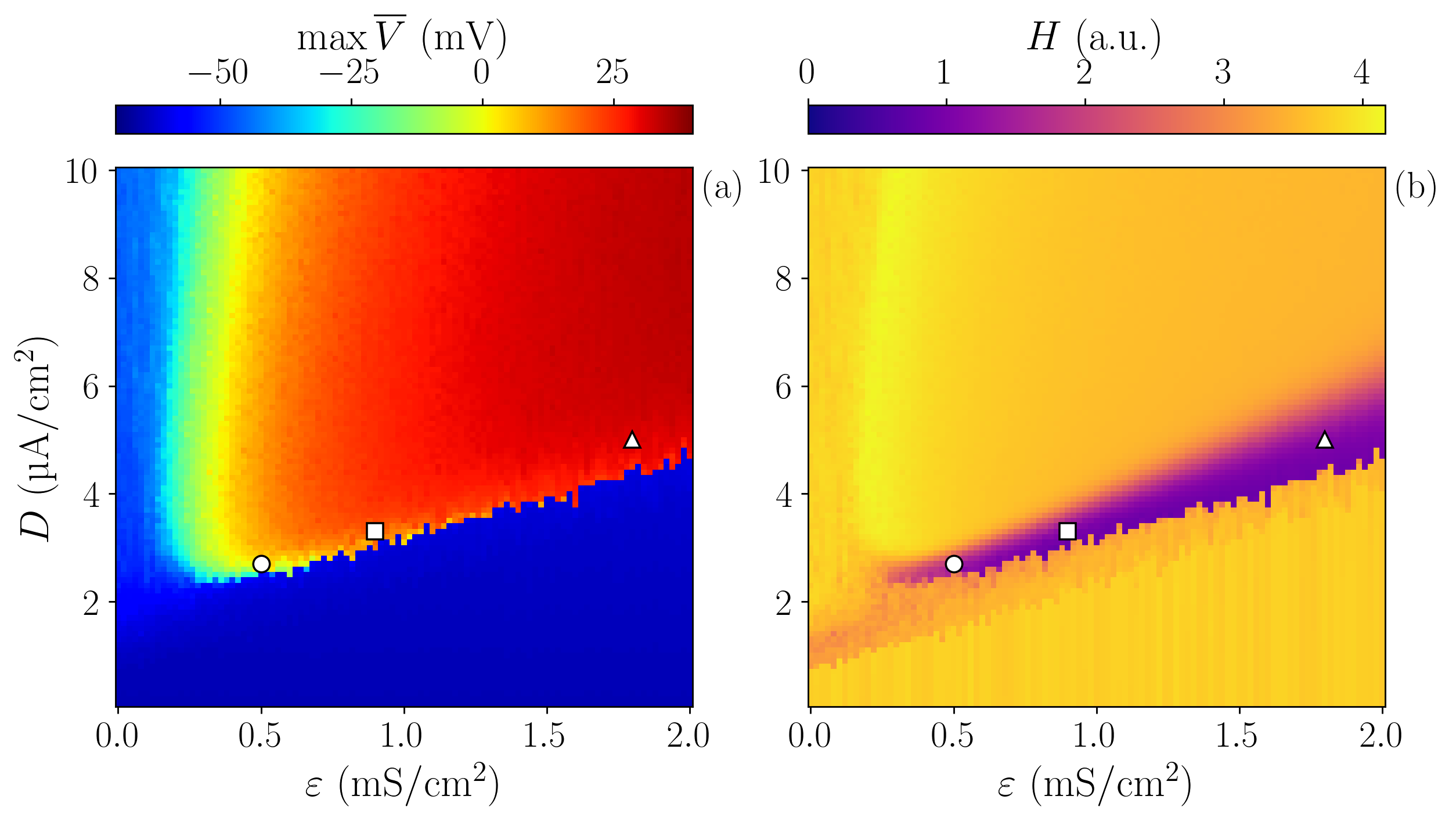}
    \caption{(a) Maxima of the mean field, $\mathrm{max}(\overline{V})$ and (b) Shannon entropy, $H$, of the distribution of mean field values, $\overline V$, in color code, in the parameter space defined by the coupling strength and the noise strength, $\varepsilon \times D$. At each point, the simulation starts from the same random initial condition.}
    \label{fig:entropy_surf}
\end{figure}

To better visualize the effects of the coupling and noise strengths, in Fig. \ref{fig:entropy_surf}(a) we show in color code the maximum of the mean field in the parameter space $\varepsilon \times D$ using a (100 × 100) grid. The same random initial condition is used for each grid point. We observe two distinct transitions from low to high $\overline V$. On the one hand, with increasing $\varepsilon$ and large enough noise ($D>2$ $\SI{}{\micro\ampere/\centi\meter^2}$) there is a smooth increase of $\overline V$; on the other hand, with increasing noise and large enough coupling (for $\varepsilon> 0.4$ $\SI{}{\milli\siemens/\centi\meter^{2}}$) there is a sudden increase of $\overline V$. 

\subsection{Entropy analysis}

To better understand the statistics of the mean field, we go beyond the analysis of its maximum value, $\mathrm{max}(\overline{V})$, and analyze the shape of the distribution of values. Specifically, we analyze Shannon's entropy,
\begin{equation}
    H = -\sum p_{i}\log p_{i},
\end{equation}
where $p_{i}$ is the probability that $\overline V(t)$ is in the i-th bin, as estimated from the histogram of $\overline V(t)$ values, we use 100 bins and adjust their range according to the minimum and maximum values of $\overline V$ for each time series. 

\begin{figure}[tb!] 
    \centering
    \includegraphics[width=.95\linewidth]{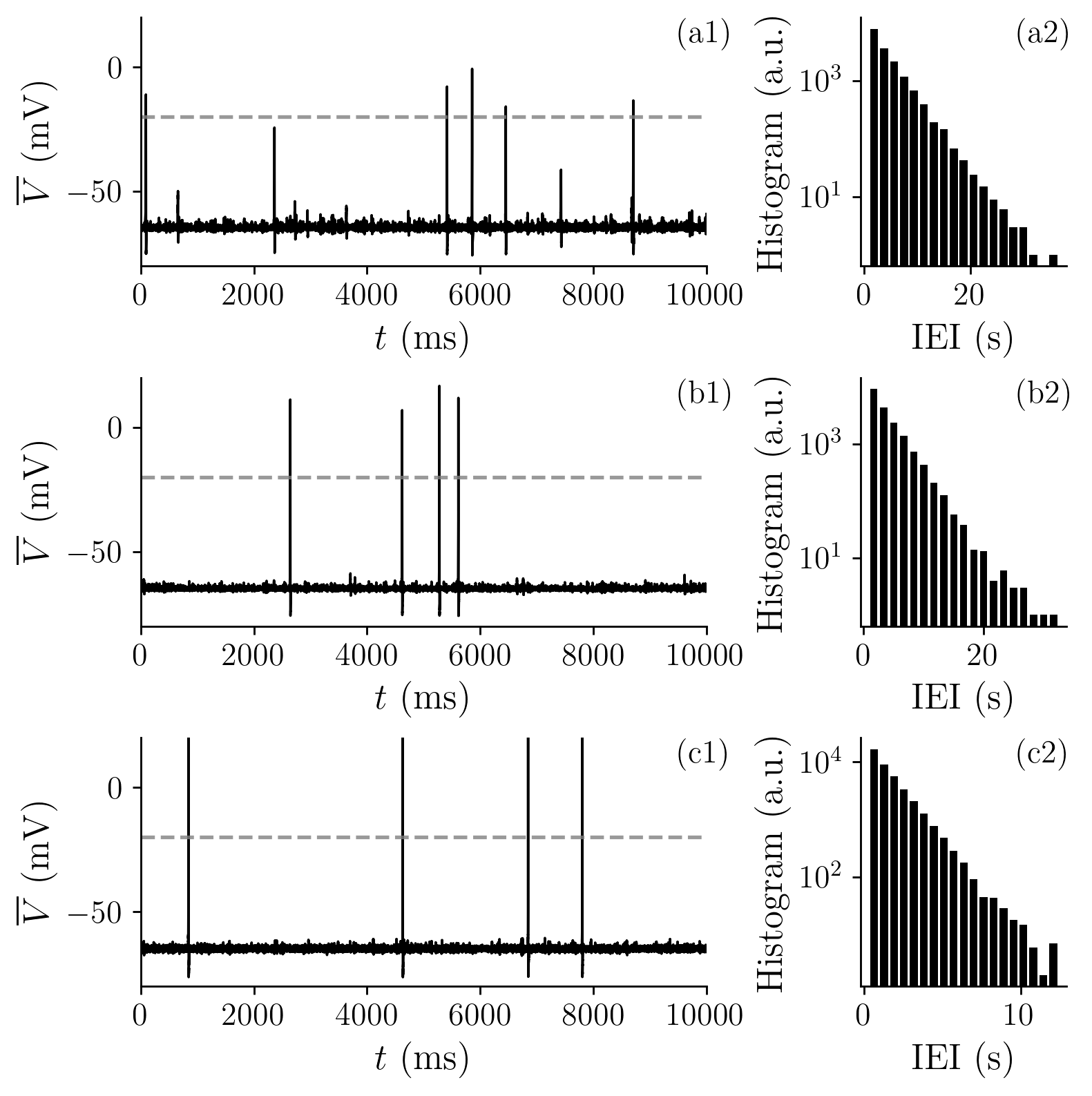}
    \caption{Time evolution of the mean field $\overline V$ (left) and the distribution of interval between extreme events (right) for $3$ distinct pairs of points in Fig. \ref{fig:entropy_surf}: (a) ($\varepsilon=0.5$, $D=2.7$) (circle), (b) ($\varepsilon=0.9$, $D=3.3$) (square), (c) ($\varepsilon=1.8$, $D=5.0$) (triangle). To compute the PDFs we evolve 100 distinct ICs for $10^6$ ($\SI{}{\milli\second}$). An extreme event is computed where $\overline V$ crosses $-20$ $\SI{}{\milli\volt}$ (gray dashed line).}
    \label{fig:events}
\end{figure}

Figure \ref{fig:entropy_surf} (b) presents $H$ for the same parameter space as panel (a), ranging from purple (low $H$) to yellow (high $H$). Most parameter space exhibits high $H$ values, but a low $H$ region can be observed in the parameter region where, in Fig. \ref{fig:entropy_surf} (a), there is an abrupt transition. Lower values of $H$ reveal less uncertainty, due to the existence of extreme values of $\overline V$: the distribution of $\overline V$ becomes narrower as it develops a tail as presented in Figs. \ref{fig:LFP_add_eps}(n) and \ref{fig:LFP_add_D}(l). This can be seen in Fig. \ref{fig:events} that displays the time evolution of the mean field, $\overline V$, as well as the distribution of values, in the three points marked with symbols in Fig. \ref{fig:entropy_surf}. We observe a decline in the variability of the extreme values of $\overline{V}$ as the parameters increase. We associate this with the requirement of higher values of $\varepsilon$ for the system to reach high $\overline V$ for stronger noise $D$. 

To complement the characterization of the mean field dynamics, we analyze the duration of the time intervals in which the network remains in the non-spiking state, until an avalanche-like extreme event occurs. We refer to these periods as inter-events-intervals (IEIs). We consider that an extreme event occurs when $\overline V$ crosses the threshold of $-20$ ($\SI{}{\milli\volt}$) (gray dashed line in Fig. \ref{fig:events}). To estimate the $\mathrm{IEI}$ distribution, we performed 100 simulations, each of $10^6$ $\SI{}{\milli\second}$, starting from random initial conditions. In Fig. \ref{fig:events} we see that the $\mathrm{IEI}$ distributions have the
exponential decay that characterizes statistically uncorrelated events.

\subsection{Role of the network size}

\begin{figure*}[tb!]
    \centering
    \includegraphics[width=0.45\linewidth]{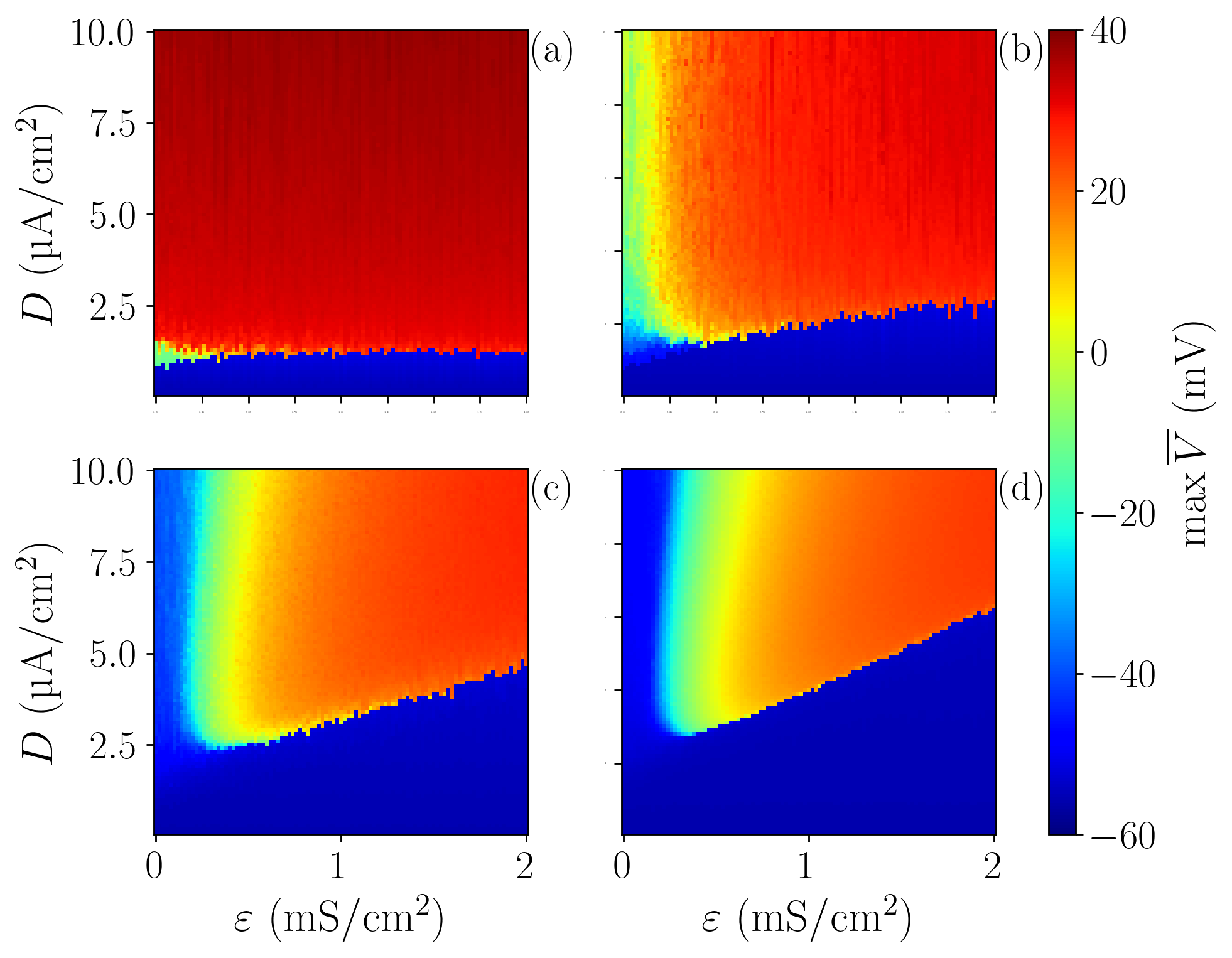}
    \includegraphics[width=0.45\linewidth]{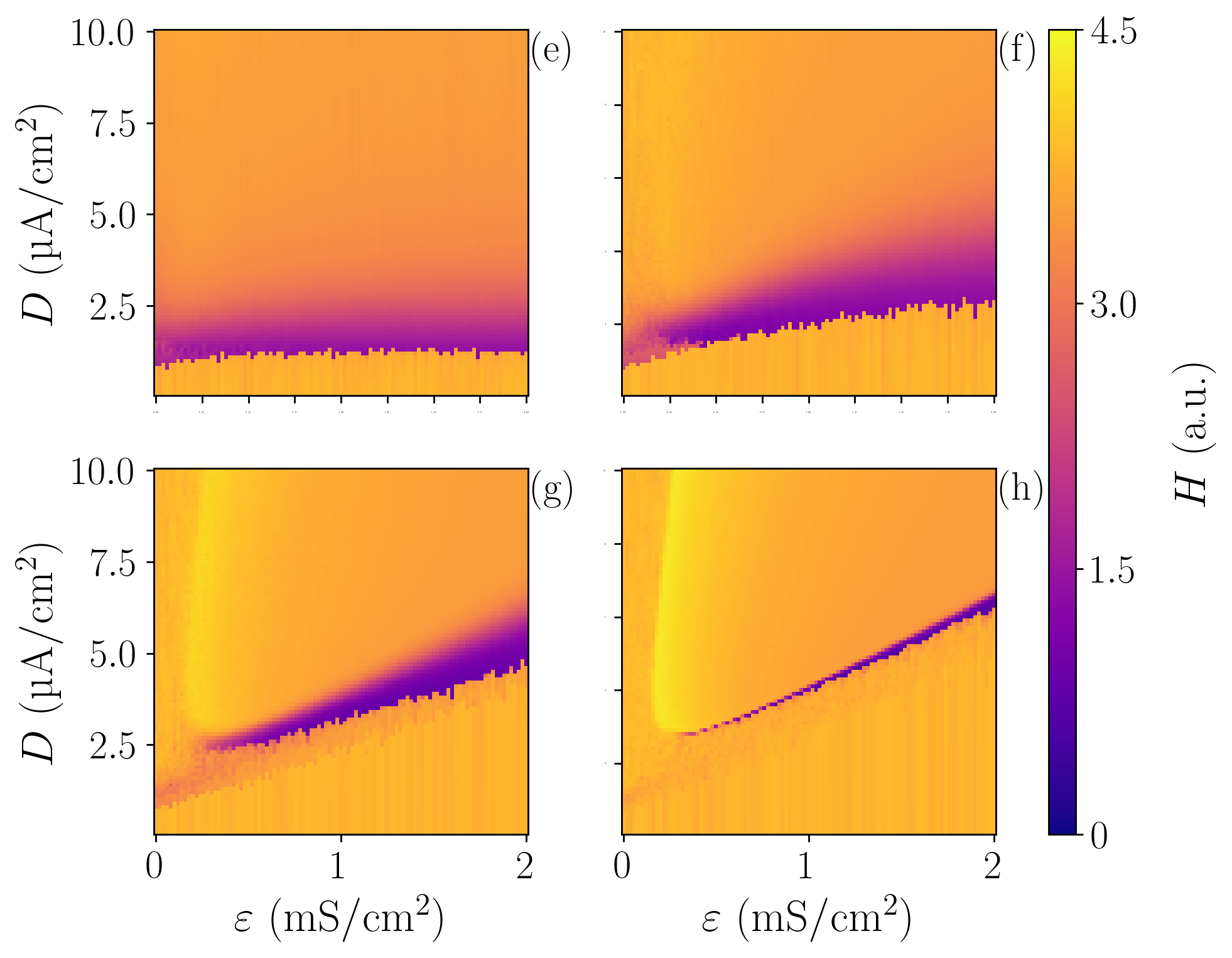}
    \caption{$\varepsilon \times D$ parameter space analysis for the maximum of the mean field ($\mathrm{max}\overline V$, left) and Shannon entropy ($H$, right) in color code for distinct network sizes: (a, e) $N=2$, (b, f) $N=10$, (c, g) $N=100$, and (d, h) $N=1000$.}
    \label{fig:size}
\end{figure*}

Next, we analyze how the number of nodes influences the overall network activity and the emergence of extreme events.  Figure \ref{fig:size} illustrates the impact of the network size by displaying the maximum value of the mean field (left) and the entropy of the distribution of mean field values (right) when the network has only two neurons, panels (a, e), 10 neurons, (b, f), 100 neurons, (c, g) and 1000 neurons, (d, h). For small networks ($N=2$ and $10$), we observe that the mean field has high amplitude values for small or even null coupling strength. As the size of the network increases, we observe low $\mathrm{max}(\overline{V})$ values for weak coupling. In addition, when increasing the network size, we observe that the region where extreme events occur becomes well-defined. This is interpreted as due to the frequency of activation of most of the neurons in the network. In small networks the neurons often become simultaneously active, leading to high values of the mean field, even for small coupling values, which are not rare events and therefore, not extremes. However, for large networks, the probability of synchronous activation decreases, and the region where extreme events occur becomes narrow and well-defined.

\section{Conclusion} \label{sec:conc}

To conclude, we have analyzed the dynamics of a network of stochastic Hodgkin-Huxley neurons with mean field coupling. We found a smooth transition from sub-threshold oscillations to spiking activity as the coupling parameter increased, followed by an abrupt suppression of spiking at higher coupling values. Similarly, increasing noise intensity led to an abrupt transition to synchronized spiking activity. These transitions highlight the dual role of coupling and noise in shaping neural activity. Specifically, while coupling initially facilitates spiking, due to its diffusive nature it ultimately suppresses it beyond a certain threshold.  By analyzing the Shannon entropy of the distribution of mean field values, we have identified parameter regions where the balance between internal (coupling) and external (noise) activity leads to the emergence of extreme events. By analyzing the maximum value of the mean field and the entropy, we have found well-defined parameter regions where these extreme events can occur.

The results demonstrate that in our model, extreme events arise from the interplay of noise and global coupling. As the size of the network increases, we observe that the region where extreme event occurs narrows. For this reason, in large neural networks, extreme events can be extremely rare. However, real neural networks are characterized by complex connectivity patterns, and ongoing work is devoted to analyzing the role of structural network connectivity on the occurrence of large-scale activity bursts. Preliminary simulations (not presented here) suggest that, for sparse complex networks, the synchronization transition for increasing coupling strength can also be abrupt. Current efforts are focused on characterizing the role of the coupling topology in the synchronization and desynchronization transitions, and in the emergence and suppression of extreme events.

For future work, it will be interesting to characterize the spiking activity of the network using the complexity-entropy plane recently used in \cite{fernanda_2021}. It will also be interesting to study spatial and temporal correlations, and compare them with the correlations occurring in real neural networks, such as those in the visual cortex of mice, recently studied in \cite{sabrina}.

Understanding which conditions can likely trigger or suppress these events is another important research question. On the other hand, we hope our findings will motivate experimental studies of networks of stochastic excitable units under global coupling. A system that could display phenomena similar to that reported here is an array of excitable lasers when it is subject to global optical or opto-electronic feedback, which will provide mean field coupling, but with a finite delay time.

\acknowledgments

B.R.R.B. and E.E.N.M. are supported by the Brazilian São Paulo Research Foundation (FAPESP), Proc. 2018/03211-6, 2021/09839-0, and 2023/16273-8. C. M. is supported by Agència de Gestió d’Ajuts Universitaris i de Recerca (2021 SGR 00606), the Institució Catalana de Recerca i Estudis Avançats (Academia), the Ministerio de Ciencia, Innovación y Universidades (PID2021-123994NB-C21) and the European Office of Aerospace Research and Development (FA8655-24-1-7022).


%

\end{document}